\documentclass{article}

\usepackage{arxiv}

\usepackage[utf8]{inputenc} 
\usepackage[T1]{fontenc}    
\usepackage{hyperref}       
\usepackage{url}            
\usepackage{booktabs}       
\usepackage{amsfonts}       
\usepackage{nicefrac}       
\usepackage{microtype}      
\usepackage{lipsum}		
\usepackage{graphicx}
\usepackage{natbib}
\usepackage{doi}
\usepackage{array}
\usepackage{tabularx}

\title{A Comprehensive Study of Supervised Machine Learning Models for Zero-Day Attack Detection: Analyzing Performance on Imbalanced Data}

\author{ Zahra Lotfi \\
	School of Electrical Engineering and Computer Science\\
	University of Ottawa\\
	Ottawa, Ontario, Canada \\
	\texttt{Zlotf035@uottawa.ca} \\
	\And
	\href{https://orcid.org/0009-0000-6663-8478}{\includegraphics[scale=0.06]{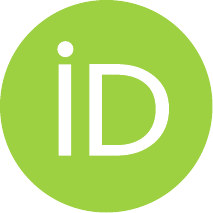}\hspace{1mm}Mostafa Lotfi} \\
	Department of Computer Engineering\\
	K.N.Toosi University\\
	Tehran, Iran \\
	\texttt{Mostafa.Lotfi@email.kntu.ac.ir} \\
}

\renewcommand{\shorttitle}{A Comprehensive Study of Supervised Machine Learning Models for Zero-Day Attack Detection}

\hypersetup{
pdftitle={A Comprehensive Study of Supervised Machine Learning Models for Zero-Day Attack Detection: Analyzing Performance on Imbalanced Data},
pdfsubject={q-bio.NC, q-bio.QM},
pdfauthor={Zahra ~Lotfi, Mostafa ~Lotfi},
pdfkeywords={Zero-Day Attacks; Class Imbalance Problem; Supervised Models; Classification Models; Oversampling; Dimension Reduction},
}

\begin{document}
\maketitle

\begin{abstract}
Among the various types of cyberattacks, identifying zero-day attacks is problematic because they are unknown to security systems as their pattern and characteristics do not match known blacklisted attacks. There are many Machine Learning (ML) models designed to analyze and detect network attacks, especially using supervised models. However, these models are designed to classify samples (normal and attacks) based on the patterns they learn during the training phase, so they perform inefficiently on unseen attacks. This research addresses this issue by evaluating five different supervised models to assess their performance and execution time in predicting zero-day attacks and find out which model performs accurately and quickly. The goal is to improve the performance of these supervised models by not only proposing a framework that applies grid search, dimensionality reduction and oversampling methods to overcome the imbalance problem, but also comparing the effectiveness of oversampling on ml model metrics, in particular the accuracy. To emulate attack detection in real life, this research applies a highly imbalanced data set and only exposes the classifiers to zero-day attacks during the testing phase, so the models are not trained to flag the zero-day attacks. Our results show that Random Forest (RF) performs best under both oversampling and non-oversampling conditions, this increased effectiveness comes at the cost of longer processing times. Therefore, we selected XG Boost (XGB) as the top model due to its fast and highly accurate performance in detecting zero-day attacks.
\end{abstract}


\section{Introduction}
Cyberattacks are increasingly posing a threat to individuals and businesses due to financial losses and data breaches. The rise of electronic services, the number of devices connected to networks, and modern technologies such as the Internet of Things (IoT) have led to the increasing prevalence of network attacks. The recent pandemic has also created a ground for attackers targeting teleworkers and healthcare facilities \cite{ref-journal1}. Despite the efforts of security organizations such as the Anti-Phishing Working Group, cyberattackers continue to evolve their methods and exploit the vulnerabilities of security mechanisms to penetrate networks and devices and misuse their information \cite{ref-journal2}. 

Some of the widely used approaches for detecting network intrusions are signature-based detection, anomaly-based detection, and heuristic-based detection systems; however, each of them has limitations in detecting network attacks, especially in the case of detecting zero-day attacks that have never been observed before and are not included in the block lists of security systems. For example, signature-based detection methods are inefficient for new or evolving attack patterns due to the lack of corresponding signatures in their signature library \cite{ref-journal3}. Zero-day attacks are sometimes used only once to steal information from specific victims and are then discarded to avoid detection by security measures. In phishing, for example, criminals generate a new Uniform Resource Locator (URL) for each attack to reduce the risk of getting caught and maximize the number of victims they can target \cite{ref-journal4}. In this case, the main challenge in zero-day phishing attacks is that the phishing URLs are created and discarded immediately after the attacker obtains the necessary information. So, they do not have time to submit a case report. In some cases, a zero-day attack is even reported to go undetected for a significant period of time (several months) before being detected by the security team or software.

To address these limitations, ML approaches and classification algorithms in particular are increasingly being used to improve the accuracy of intrusion detection systems. Some of these are described in the literature review section. However, these models are vulnerable to identifying the minority class of actual cyberattacks in a significant amount of normal network data. This challenge, known as the class imbalance problem, presents a training obstacle for models, as they often struggle to correctly classify the rare attacks into the vast majority of legitimate traffic. Consequently, significantly imbalanced data sets pose a frequent challenge for machine learning methods due to the model's ability to generalize to the minority class \cite{ref-journal5}.

Taking into account the mentioned factors, we developed a framework to address the class imbalance problem and study the performance of multiple supervised models in detecting zero-day attacks. These models are evaluated on a comprehensive benchmark data set containing 2,218,761 normal and 321,283 attack data sets, which makes them highly unbalanced. The data set contains a variety of features extracted from packet flows. During the selection of the data set and classification models, we considered the following crucial factors as in \cite{ref-journal6}: 

\begin{enumerate}
\item	Selecting appropriate and effective ML models,

\item   Leveraging special and informative features,

\item   And collecting a comprehensive set of representative samples to train the model.
\end{enumerate}

The research has three primary aims: i) investigate the effectiveness of various lightweight binary classification techniques on a specific benchmark data set, particularly in identifying zero-day attacks; ii) evaluate the influence of oversampling in addressing the data set's imbalance issue; iii) and assess the execution time of each model to determine the feasibility of creating a high-performance yet fast model. The rationale behind the last objective is the distinction between merely identifying attacks and doing so efficiently. An intrusion detection system loses its efficacy if it takes an extended time to process incoming data and detect attacks while it's distributed across a network, or halts traffic until classification is complete. The paper is organized as follows: Section ~\ref{sec 2} reviews the existing literature related to the application of ML techniques for recognizing network attacks and URL phishing. Section ~\ref{sec 3} outlines the framework designed for the study, including its development and implementation. Section ~\ref{sec 4} presents the implementation outcomes. Section ~\ref{sec 5} delves into the research findings, and Section ~\ref{sec 6} concludes the paper.

\section{Related Work}
\label{sec 2}

Although there are a variety of network security approaches, users continue to be affected by network attacks. Over the last ten years, ML models have been widely applied in detecting network intrusions; some include zero-day detection techniques. Table ~\ref{tab1} provides a summary of the latest articles (published between 2018 and 2023) analyzed in this study, indicating the ML methods and data sets used in these articles. The complete model names are provided below the table.

\begin{table*}
\renewcommand{\arraystretch}{1.4}  

\caption{Summaries of reviewed papers applying ML models in network intrusion and URL phishing detection\label{tab1}}

\begin{tabularx}{\textwidth}{%
>{\centering\arraybackslash}m{3.5cm}
>{\centering\arraybackslash}m{4cm}
>{\centering\arraybackslash}m{1.5cm}
>{\centering\arraybackslash}m{1cm}
>{\centering\arraybackslash}m{4.5cm}
}
\toprule
\textit{Reference} & \textit{ML Methods} & \textit{Type of Data} & \textit{Zero Day?} & \textit{Data Set Source} \\
\midrule
{\cite{ref-journal7}} & MLP*, RF* & Text & No & PhishTank, Tranco \\
{\cite{ref-journal8}} & MLP*, RF* & PCAP & Yes & UNSW-NB15, NF-UNSW-NB15-v2 \\
{\cite{ref-journal9}} & DL*, GAN* & Text & Yes & Kaggle Website \\
{\cite{ref-journal3}} & One-class SVM*, AutoEncoder & PCAP & Yes & CSE-CIC-IDS2018, NSL-KDD, CIC-IDS2017 \\
{\cite{ref-journal10}} & Supervised Learning & PCAP & Yes & UNSW-NB15, BoT-IoT, ToN-IoT, CSE-CIC-IDS2018 \\
{\cite{ref-journal11}} & Graph CNN* + LRCN* & Text & Yes & Google Search Engine, PhishTank \\
{\cite{ref-journal1}} & MLP + TF-IDF & Text & No & Not Specified \\
{\cite{ref-journal12}} & SVM, XGB*, RF & PCAP, CSV & Yes & KDD99 and NSL-KDD, CIC-IDS2018 \\
{\cite{ref-journal6}} & NLP Feature Extraction + RF, DT*, LR*, NB* & Text & No & Alexa and Yandex, PhishTank and OpenPhish \\
{\cite{ref-journal4}} & CAE* & Text & Yes & ISCX-URL-2016, PhishTank, PhishStorm \\
{\cite{ref-journal13}} & CNN & Text & Yes & PhishTank, Common Crawl Foundation \\
{\cite{ref-journal14}} & CNN + Similarity Metrics & Visual Data & No & PhishTank \\
{\cite{ref-journal2}} & RF, KNN, DT, SVM, LR & Text, Structured Data & No & Not Specified \\
{\cite{ref-journal15}} & Semantic Feature Extraction + RF, LR & Text & Yes & Ebbu2017, DMOZ, Alexa.com \\
{\cite{ref-journal16}} & NLP* Feature Extraction + DT, Adaboost, K-star, KNN, RF, SMO, NB & Text & Yes & Ebbu2017 \\
\toprule
\end{tabularx}
\end{table*}

As can be seen, all studies used supervised classification models, most of them (11 out of 15 papers) used Neural Networks (NN) and deep learning in their models. In addition to a study \cite{ref-journal13} that used visual data in their model, other works also used URLs and PCAP files in their work, and some of them used NLP-based feature extraction methods to extract structured data extract \cite{ref-journal11, ref-journal14, ref-journal15}. The table also shows that 10 out of 15 works are examined for the detection of zero-day attacks.

Some studies \cite{ref-journal8, ref-journal10} present the results of ML accuracy for different types of attacks in tabular form. In particular, some attack types have lower accuracy compared to others because of the lack of suitable samples. \cite{ref-journal8} presents a novel application of Zero-Shot Learning (ZSL) within Network Intrusion Detection Systems (NIDS) to address the challenge of detecting novel cyberattacks, particularly zero-day network intrusions. By leveraging ZSL, the authors propose a method for training ML models using known attack data. These models can then extract and learn characteristic features (semantic attributes) that define different attack behaviors. During testing, the models use these learned attributes to classify entirely new classes of attacks, including previously unknown zero-day attacks. This data set is used in our research to apply multiple classifiers (RF, DT, LR, MLP, and XGB) along with dimensionality reduction and oversampling techniques to examine their performance in identifying zero-day attacks. Our goal is also to determine whether optimizing the hyper-parameters of the given models leads to a higher performance score.

Among the articles discussed, \cite{ref-journal12} addresses the class imbalance problem and explores the use of ML to build an intrusion detection system (IDS) that can identify known and zero-day cyberattacks. To address this issue and improve the effectiveness of the ML models, the article explores under/oversampling techniques as methods to balance the data set. This approach ensures that models are trained on a more representative distribution of network traffic patterns, ultimately leading to better detection of malicious activity. 

In detecting phishing attacks, \cite{ref-journal14} introduces VisualPhishNet, a similarity-based phishing detection framework that applies a CNN model. Using a similarity metric, this framework detects phishing websites, especially on pages with a new look. The authors also created the largest visual phishing detection data set by searching active phishing pages on the PhishTank website and collecting screenshots from the pages, resulting in 10,250 examples. The evaluation metrics they considered when detecting phishing samples were the size, color, location, and website languages of the elements. 

CNN is also used in other research to detect phishing websites solely by analyzing URLs \cite{ref-journal13}. The authors explain that unlike other studies that split URLs into different parts and analyze each part to identify phishing attacks, their model only needs to encode URLs as character-level one-hot vectors. The vector is then passed to the CNN and the model can detect phishing samples with almost 100\% accuracy. It is also suitable for detecting zero-day attacks.

PhishDet is another neural network-based framework for detecting zero-day phishing attacks. Developed based on Graph Convolutional Neural Networks (GCNN) and LRCN, this model can detect malicious websites by analyzing their URLs and HTML codes with an accuracy of 96.42\%. To maintain its high performance, PhishDet requires frequent retraining over time \cite{ref-journal11}. Using a CAE \cite{ref-journal14} provides character-level URL functional modeling to detect zero-day phishing attacks. They used three real-world phishing data sets containing 222,541 URLs to examine their model and considered Receiver-Operating Characteristic (ROC) as a metric to compare their results with the results of other models in the literature.

Regarding non-NN models, four articles were written to study classification techniques. \cite{ref-journal2} proposes an ML-based classification algorithm that uses heuristic features such as URL, source code, session, security type, protocol and website type to detect phishing websites. The algorithm is evaluated using five ML models. The RF algorithm is considered the most effective, achieving an attack detection accuracy of 91.4\%. Furthermore, \cite{ref-journal16} propose a real-time anti-phishing system that uses seven different classification algorithms and NLP-based functions. The system differs from other literature studies in its language independence, use of a large data set, execution in real time, detection of new websites and use of feature-rich classifiers. The experimental results show that the RF algorithm with only NLP-based features achieved the best performance in detecting phishing URLs.

One of the studies in which non-NN models have been applied is conducted with the main goal of creating a balanced benchmark data set containing an equal number of harmless and malicious URLs \cite{ref-journal6}. The collected URLs are processed and 87 features are extracted from URLs, web page content and external resources retrieved through third-party queries (such as Alexa and WHOIS). The authors then applied five different classification methods (RF, DT, LR, NB, and SVM) to see their performance on the created data set. However, they did not consider the issue of zero-day attacks in their work.

\section{Materials and Methods}
\label{sec 3}

This section describes the zero-day attack detection framework with details of the data set and the ML models used in this work. The proposed framework aims to study the performance of several supervised algorithms on a popular Netflow data set to evaluate the models' performance in detecting zero-day attacks. Two main differences characterize this framework: i) the application of a significantly imbalanced data set and ii) the implementation of zero-day attacks in a way that remains unknown to the classification models in order to emulate the real situation.

Figure 1 shows how the framework functions and how zero-day attacks are implemented in it. This framework consists of three main components. The first component prepares the zero-day attack set and inserts it into the test set in a random distribution to prevent the model from being biased by the location of the zero-days in the data set. The second component aims to prepare the train and test set through preprocessing, standardization, dimensionality reduction to select the most relevant features, and oversampling to solve the class imbalance problem. The third component performs hyper-parameter tuning, training and evaluation of the candidate classification models to select the champion model with the highest performance and reasonable running time.

\begin{figure*}
    \centering
    \includegraphics[width=\textwidth]{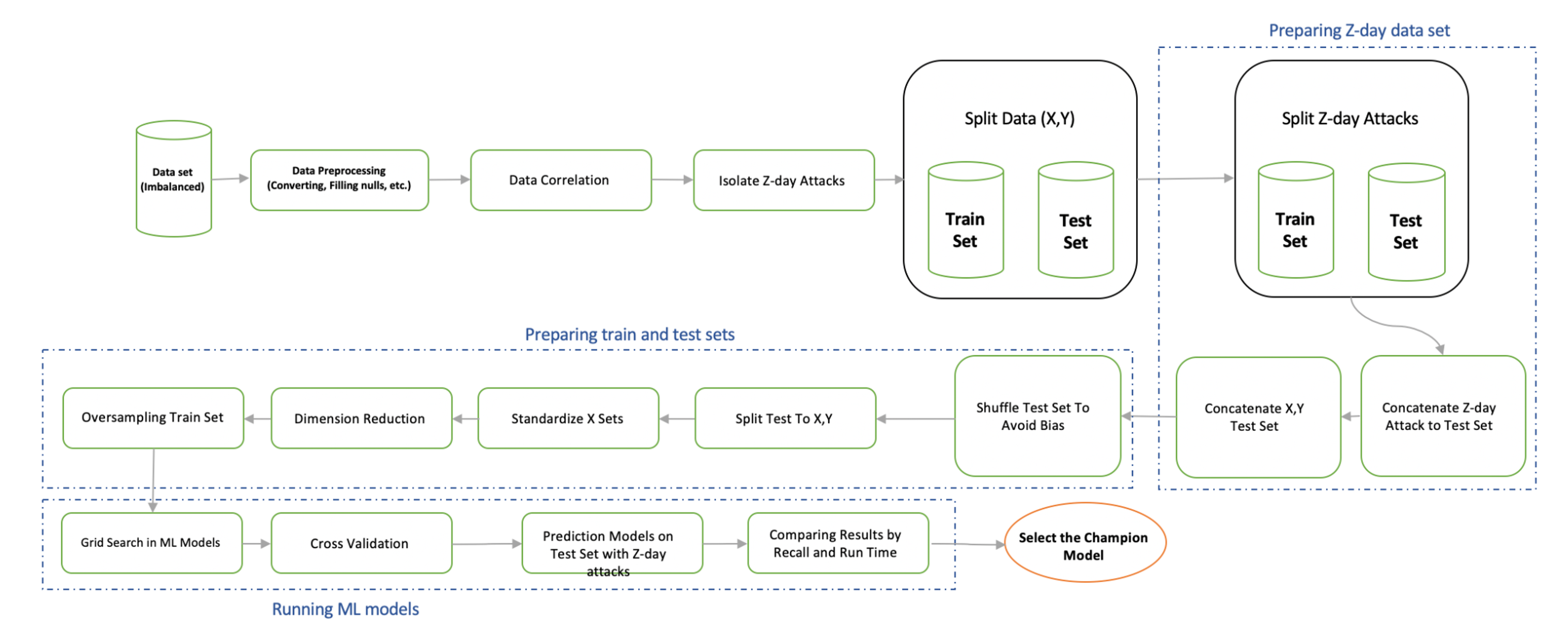}
    \caption{Diagram of the proposed URL phishing detection framework}
    \label{fig1}
\end{figure*}

As shown in Figure 1, most of the tasks performed in this framework are related to data preprocessing and feature engineering because the implementation of zero-day attacks is important. This type of attack should be isolated from ML models in the training phase and then presented in the testing phase. It is therefore important to implement the model in such a way that it reflects the real situation. To do this, we first split the data set into training and test sets (by 70\% to 30\%) and then added a zero-day attack set to the test set. The “shuffling” step plays a key role in randomizing the test data set. Otherwise, if the zero-day data set was appended to the test set, the data set would be biased, allowing models to falsely detect the attacks with significantly higher performance, as shown in Table ~\ref{tab2}.

\begin{table*}[ht]
\centering
\caption{The results of predictions on biased zero-day attack detection\label{tab2}}
\begin{tabular*}{\textwidth}{@{\extracolsep{\fill}} >{\centering\arraybackslash}c c c c c c c@{}}
\toprule
\textit{ML Model} & \textit{Accuracy} & \textit{Recall} & \textit{Precision} & \textit{F1 Score} & \textit{ROC AUC} & \textit{FPR} \\
\hline
LR    & 98.91\% & 96.45\% & 93.12\% & 94.77\% & 98.08\% & 94.40\% \\
DT    & 99.33\% & 95.16\% & 92.04\% & 93.56\% & 97.35\% & 95.25\% \\
RF    & 99.42\% & 96.44\% & 93.11\% & 94.72\% & 98.88\% & 96.46\% \\
MLP   & 99.39\% & 96.17\% & 94.08\% & 93.89\% & 97.93\% & 94.10\% \\
XGB   & 99.35\% & 95.62\% & 92.66\% & 94.18\% & 96.67\% & 95.49\% \\
\toprule
\end{tabular*}
\end{table*}

Before delving into the details of the framework, the next section describes the selected data set and the process of preparing the train and test sets.

\subsection{Data Set}
\label{subsec 1}

UNSW-NB15, a popular open source data set for evaluating NIDS, is used. With 49 fields and 2,540,044 records, including 2,218,761 benign and 321,283 attack records, this data set is accessible on Kaggle. The total number of attacks for each type is shown in Table ~\ref{tab3}. Two fields are used to label attacks: "Label" indicates whether the example is a normal (0) or an attack (1), and "attack-cat" indicates the type of attack. 

\begin{table*}[ht]
\centering
\caption{The total number of benign (normal) and attack records in the UNSW-NB15 data set.\label{tab3}}
\begin{tabular*}{\textwidth}{@{\extracolsep{\fill}} >{\centering\arraybackslash}c c c c c c@{}}
\toprule
\textit{Type} & \textit{Normal} & \textit{Generic} & \textit{Exploits} & \textit{Fuzzers} & \textit{DoS} \\
\hline
No. Records & 2,218,761 & 215,481 & 44,525 & 24,246 & 16,353 \\
Type & Reconnaissance & Analysis & Backdoors & Shellcode & Worms \\
No. Records & 13,987 & 2,677 & 2,329 & 1,511 & 174 \\
\toprule
\end{tabular*}
\end{table*}

To create the data set, the tcpdump tool is used to capture 100GB of network packets, which are stored in 4 different CSV files to facilitate downloading and working with data. Since the entire data set is very large and requires huge computing and storage resources, half of it is selected for this research, which contains 1,400,002 data sets, including 1,325,038 benign and 75,063 attack examples. Table ~\ref{tab4} shows the total number of attacks per type in the selected data set.

\begin{table}
\centering
\caption{The total number of benign (normal) and attack records in the UNSW-NB15 data set\label{tab4}}
\begin{tabular*}{\columnwidth}{@{\extracolsep{\fill}}c c c@{}}
\toprule
\textit{Sample Type}	& \textit{Number of Records}	& \textit{Percentage}\\
\hline
Generic	        & 35,405		& 47.23\%\\
Exploits	    & 16,512		& 22.03\%\\
Fuzzers		    & 9,719			& 12.97\%\\
DoS             & 5,804			& 7.74\%\\
Reconnaissance	& 4,875			& 6.50\%\\
Analysis		& 1,134			& 1.51\%\\
Backdoor        & 904		    & 1.21\%\\
Shellcode       & 547			& 0.73\%\\
Worms	        & 64			& 0.08\%\\
\hline
\textit{Total Attacks}	& \textit{74,964}			& \textit{5.36\%}\\
\textit{Normal Samples}	& \textit{1,325,038}		& \textit{94.64\%}\\
\textit{Total Samples}	& \textit{1,400,002}		& \textit{100\%}\\
\toprule
\end{tabular*}
\end{table}

From the percentage of Total Attacks, it is obvious that the data set is highly imbalanced and causes the class imbalance problem in classification models. It can also be seen that the frequency of attacks varies and some are significantly smaller than others. 

To prepare the data set for zero-day attacks, based on the approach proposed in \cite{ref-journal12}, a group of attacks is selected, isolated from the training data set, and then inserted into the test set including newly collected and unseen samples, the number of which is smaller compared to known attacks. The selected group consists of four least common attack types, including worms, shellcode, backdoor, and analysis. Therefore, the framework is designed to predict invisible zero-day attacks among other trained attacks by using the binary field “Label” that specifies normal and attack examples.

\subsection{Data Preprocessing}
\label{subsec 2}

UNSW-NB15 is a relatively clean data set, but some types of cleansing and preprocessing were required to prepare the data for training and prediction. For the attack-cat field, non-attack records (with label = 0) were null and replaced with Normal. Most fields were numeric, but there were some object-like fields that converted to numeric values. The attack-cat field also required minor changes to standardize all category names. After managing the data type of the features, the null values, and the mismatched category labels, the next step was to examine the correlation between the features using a heat map chart (Figure 2). The features were then sorted according to their correlation with the target fields (label). This step resulted in the removal of the less relevant and less efficient features from the data set. 

\begin{figure}[ht]
    \centering
    \includegraphics[width=\columnwidth]{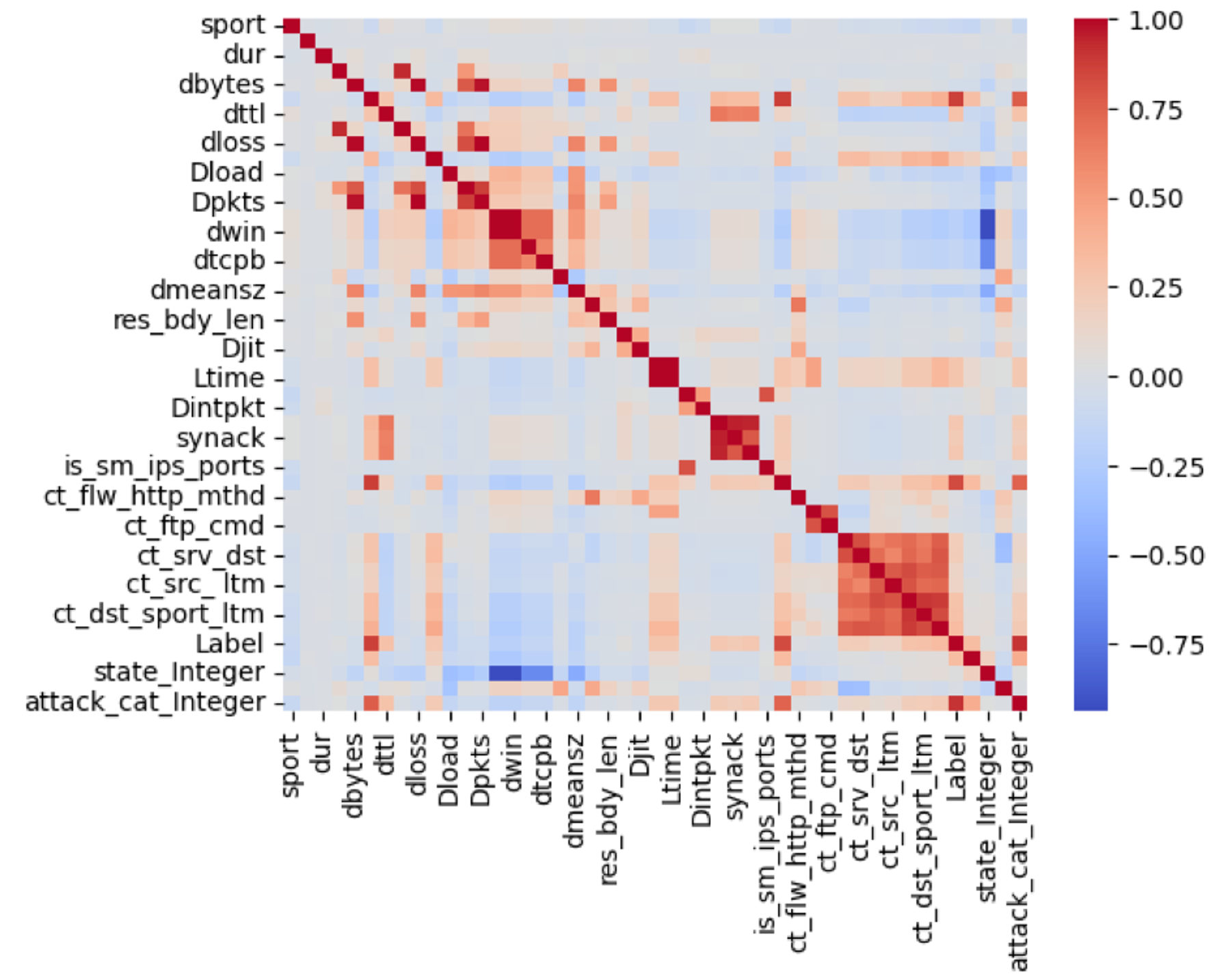}
    \caption{The heat map of features’ correlations}
    \label{fig2}
\end{figure}

The final step of preprocessing is to apply dimensionality reduction methods to the selected data set to make it suitable for ML models. To do this, all values of the original data set (X set) converted into numbers should be normalized and transformed on a similar scale and made comparable using the dimension reduction process. In addition, standardization leads to improvement in the performance of ML algorithms.

\subsection{Dimension Reduction and Resampling}
\label{subsec 3}

There are various methods to reduce the dimension of a data set. One of the most popular is Principal Component Analysis (PCA). PCA analyzes complex data sets with multiple variables to extract important information while minimizing the loss of relevant data by reducing dimensionality and retaining the maximum possible variation \cite{ref-book1}. The reason why PCA was chosen for this research is that PCA is not sensitive to noise, requires less memory and computational units, and is sufficiently efficient when working on data sets that are not very high dimensional \cite{ref-journal17}. . Due to the large size of our data set and the lack of hardware capacity, PCA was the best choice to reduce the size of the data set.

Applying PCA to the data set and then examining the output of the classification models shows how relevant and important the removed features are for prediction. To find the best number of components to use in PCA, we gave PCA a range of (1, data set.Shape + 1) and graphed the results to find the number of components that result in a variance threshold of more than 95\%. The results show that 24 components should be selected for the X set (Figure 3).

\begin{figure}[ht]
    \centering
    \includegraphics[width=\columnwidth]{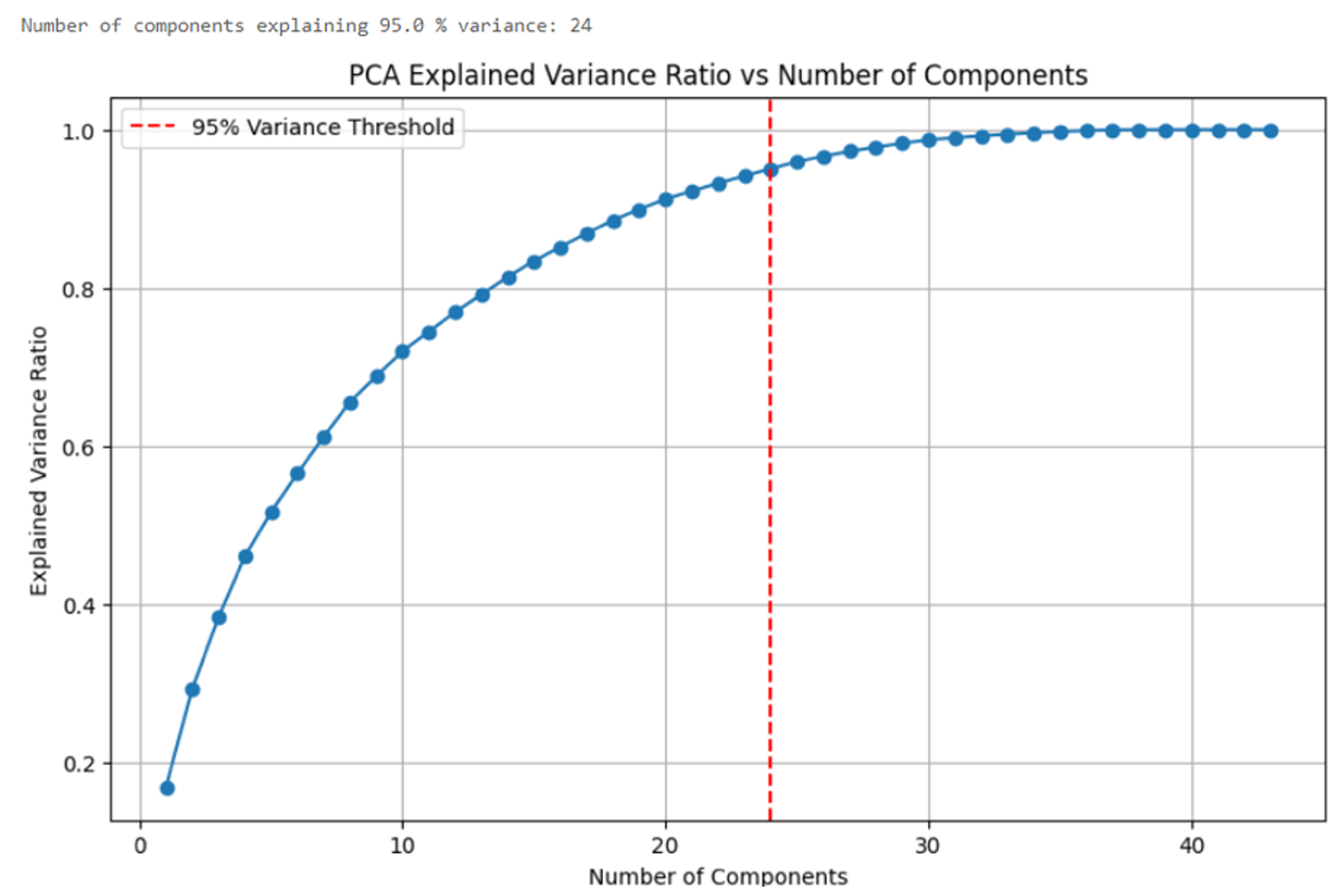}
    \caption{Finding the best number of components for PCA}
    \label{fig3}
\end{figure}

As mentioned previously, the UNSW-NB15 data set is highly imbalanced, which means that it includes minority classes with significantly fewer samples compared to the majority class. This problem causes the classification models to be inefficient at detecting attacks, despite having considerably high accuracy (Table ~\ref{tab5}).

\begin{table*}[ht]
\centering
\caption{The performance of classification models in an imbalanced data set\label{tab5}}
\begin{tabular*}{\textwidth}{@{\extracolsep{\fill}}c c c c c@{}}
\toprule
\textit{ML Model} & \textit{Accuracy} & \textit{Recall} & \textit{Precision} & \textit{F1 Score} \\
\hline
LR  & 97.41\% & 37.56\% & 34.56\% & 35.86\% \\
DT  & 97.67\% & 33.50\% & 32.55\% & 31.83\% \\
RF  & 97.98\% & 36.40\% & 37.28\% & 35.46\% \\
MLP & 98.05\% & 39.65\% & 39.76\% & 37.26\% \\
XGB & 97.82\% & 36.66\% & 36.25\% & 35.17\% \\
\toprule
\end{tabular*}
\end{table*}

To address this issue, resampling techniques should be applied to balance the proportion of anomalies compared to normal samples in the training set. In this research, the Synthetic Minority Over-sampling Technique (SMOTE) addresses the imbalance problem by creating new data points in a similar path to existing ones. This is done by interpolating between existing data points for minority classes, resulting in a more balanced representation of the training data. This allows the model to learn the characteristics of the minority class more effectively and potentially improve classification performance for this underrepresented group \cite{ref-journal19}. The following lines show the number of attacks (1s) added to the training set after performing the SMOTE oversampling technique:

\begin{verbatim}
Number of 0s in not-oversampled training y:
927641
Number of 0s in oversampled training y: 
927641.
Number of 1s in not-oversampled training y: 
50506.
Number of 1s in oversampled training y: 
927641.
\end{verbatim}

\subsection{Grid Search}
\label{subsec 4}

Each classification model used in our framework has several hyper-parameters (or Hparams for short) that affect the process and time of learning in the training phase. To achieve optimal performance in the classification models, Hparams should be tuned with the most likely values. Various methods can be used to optimize Hparams, including grid search.

Grid search is one of the simplest search algorithms and provides extremely accurate predictions if sufficient resources are available. This approach allows users to consistently identify the optimal parameter combinations \cite{ref-journal20}. In contrast to Bayesian optimization and Bayesian search, grid search can be carried out in parallel relatively easily because each trial runs independently and is not influenced by timing. This flexibility in resource allocation makes grid search particularly beneficial for distributed systems. 

A comparative analysis by \cite{ref-journal21} highlights the superior performance of grid search over random search, regardless of whether oversampling and normalization techniques are applied. Still, grid search is not without its challenges. It is significantly affected by the curse of dimensionality, which leads to an exponential increase in resource consumption as the number of Hparams to be optimized grows \cite{ref-journal20}. To address this issue, we used principal component analysis (PCA), a well-established dimensionality reduction method.

Grid search is a method for generating different model configurations by analyzing a range of values for each Hparam of interest. The approach involves training and testing models across all combinations of values for all Hparams. Although the technique is easy to use, it can be expensive due to the large number of Hparams and levels of each one and as a result, the computational cost increases exponentially \cite{ref-journal22}. To use grid search, the Hparams are given a range of values as shown in Figure 4.

\begin{figure*}
    \centering
    \includegraphics[width=\textwidth]{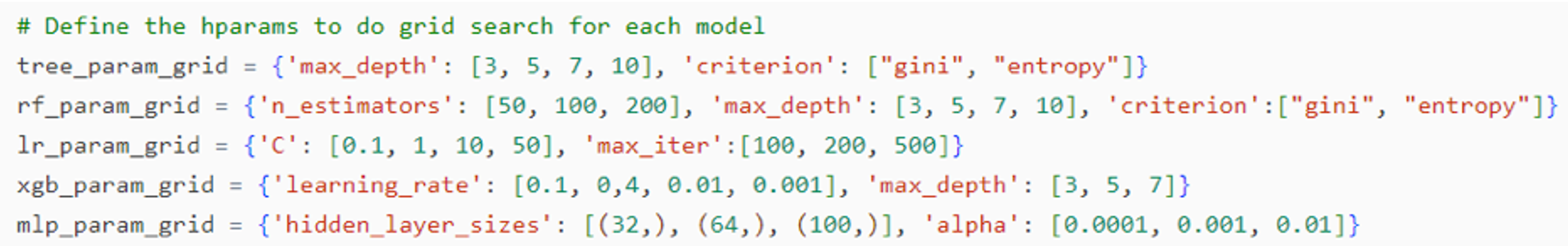}
    \caption{Values for Hparam tuning}
    \label{fig4}
\end{figure*}

Based on the results obtained from Grid Search, the classification models are set with the resulting values, shown Below:

\newpage

\begin{verbatim}
Performing grid search for Decision Tree
Best parameters: 
{'criterion': 'entropy', 'max_depth': 10}
Best score: 0.996

Performing grid search for Logistic Regression
Best parameters: 
{'C': 0.1, 'max_iter': 200}
Best score: 0.995

Performing grid search for XGBoost
Best parameters: 
{'learning_rate': 0.1, 'max_depth': 7}
Best score: 0.996

Performing grid search for Random Forest
Best parameters: 
{'criterion': 'entropy', 'max_depth': 10, 
'n_estimators': 200} 
Best score: 0.996

Performing grid search for MLP
Best parameters: 
{'alpha': 0.0001, 'hidden_layer_sizes': (32,)}
Best score: 0.996
\end{verbatim}

\subsection{Classification Models}
\label{subsec 5}

The data set used is labeled, with each entry indicating characteristics of a legitimate (normal) instance or attack. Essentially, supervised machine learning models facilitate the ability of computers to identify patterns derived from labeled input data and subsequently enable them to predict outcomes (labels) for previously unobserved samples. Therefore, supervised models represent a very suitable option.

In this framework, five different supervised models are selected to represent different approaches to classification as follows: DT and LR as two widely used instances of ML algorithms, RF and XGB as representatives of ensemble learning and MLP for deep learning approaches. Here are brief definitions for each of these models and approaches: DT is a popular ML model that predicts outcomes by recursively dividing the feature space into subspaces that serve as the basis for prediction \cite{ref-journal23}. LR is another one that is often used for classification, especially cyberattack prediction, due to its binary nature \cite{ref-journal17}. 

Ensemble learning technique is a subset of ML that combines multiple ML models to improve prediction performance. RF, introduced by Breiman  \cite{ref-journal24}, is an ensemble method for classification that creates multiple DTs to partition the data and predict classes by majority vote. XGB is another popular ensemble method based on a Gradient Boosting Machine (GBM) model and is used for both classification and regression problems. It is popular among data scientists due to its ability to perform high-speed computations outside of the core \cite{ref-journal25}.

Last but not least, MLP is a type of deep learning technique that applies artificial neural networks, which typically consist of three or more layers of nodes. MLP uses backpropagation as a supervised learning technique for training and can identify data that is not linearly separable, making it a suitable choice for binary and multiclass classifications \cite{ref-journal26}. Selecting different classification models from different learning techniques prevents our research from focusing on a specific technique and helps evaluate different supervised algorithms on a similar data set.

To predict the target class label, classification models should first be trained on a good number of samples and preferably in different rounds of execution. This can be achieved through cross-validation as an effective technique to ensure the robustness and generalizability of predictive models. By strategically splitting the training data into different folds, cross-validation makes it easier to evaluate a model's performance based on unseen information. This approach overcomes the limitations associated with evaluating a model based solely on the available data on which it was trained, thereby reducing the risk of overfitting. Table ~\ref{tab6} shows the incredible performance of ML models when performing cross-validation on the training set (after performing oversampling). The recall, F1 score, and precision metrics show that the models were able to accurately detect almost all normal and attack patterns. The next section explains how to evaluate the proposed model and what the prediction results show.

\begin{table*}[ht]
    \caption{The result of cross validation.}
    \label{tab6}
    \centering
    \begin{tabular*}{\textwidth}{@{\extracolsep{\fill}}c c c c c c c@{}}
        \toprule
        \textit{ML Model} & \textit{Accuracy} & \textit{Recall} & \textit{Precision} & \textit{F1 Score} & \textit{ROC AUC} & \textit{Run Time} \\
        \hline
        LR   & 99.59\% & 99.90\% & 99.05\% & 99.55\% & 99.62\% & 42.61 \\
        DT   & 99.58\% & 99.92\% & 99.20\% & 99.62\% & 99.88\% & 190.41 \\
        RF   & 99.68\% & 99.98\% & 99.25\% & 99.95\% & 99.96\% & 5622.65 \\
        MLP  & 99.63\% & 99.90\% & 99.12\% & 99.89\% & 99.91\% & 374.29 \\
        XGB  & 99.35\% & 99.91\% & 99.36\% & 99.78\% & 99.93\% & 20.21 \\
        \toprule
    \end{tabular*}
\end{table*}

\section{Results}
\label{sec 4}

The next step after training the model using the cross-validation technique is to examine them against the test set, including the zero-day attack set that the models are not exposed to. To evaluate the proposed framework, various metrics are considered, including recall, which quantifies the proportion of true positive cases correctly identified by a model \cite{ref-journal27}; Precision, which reflects the model's ability to accurately identify true positives \cite{ref-journal28}; F1 score, which eliminates the limitations of relying solely on precision or recall by providing a harmonious mean between the two \cite{ref-journal27}; False Positive Rate (FPR), i.e. the proportion of negative cases that were incorrectly classified as positive by the model \cite{ref-journal28}; AUC ROC, which quantifies the performance of a classification model at different classification thresholds. It reflects the model's ability to distinguish between positive and negative cases \cite{ref-journal28}; And accuracy, which represents the overall proportion of correctly classified instances within the data set \cite{ref-journal27}. 

Unlike other studies that do not consider the time it takes for models to classify the test set, we considered the running time as an evaluation metric, not for the model's prediction performance, but as a measure of whether the model is a suitable candidate in intrusion Detection systems can be used without the detection of attacks taking a lot of time. It is worth noting that to identify anomalies (in our research attacks), Recall, Precision and F1-Score are mostly used instead of Accuracy. This is because the accuracy is usually high even if the model cannot accurately classify attacks due to the high number of normal cases, while other metrics mentioned show the performance of the models based on the number of attacks correctly labeled.

In our framework, the number of zero-day attacks in the test set is significantly lower than other attacks and normal samples (only 0.63\%). Therefore, the performance of the models is expected to decrease dramatically. Hence, it is important to apply methods to improve prediction performance. Thus, in addition to techniques such as standardization, dimensionality reduction, and cross-validation, we equipped the framework with the oversampling method (SMOTE) to balance the train set. To determine how oversampling affects prediction performance, we applied two parallel approaches when running the classification models: one without oversampling and one with oversampling. Table ~\ref{tab7} shows how the classification models affect the sentence before the SMOTE method is applied. As shown in the table, running time is also added to the evaluation metrics.

\begin{table*}
\caption{Classification models' performance before oversampling technique\label{tab7}}
\begin{tabular*}{\textwidth}{@{\extracolsep{\fill}}c c c c c c c c@{}}
\toprule
\textit{ML Model} & \textit{Accuracy} & \textit{Recall} & \textit{Precision} & \textit{F1 Score} & \textit{ROC AUC} & \textit{FPR} & \textit{Run Time} \\
\hline
{LR} & 96.14\% & 45.00\% & 79.48\% & 57.49\% & 72.10\% & 43.67\% & 9.98 \\
{DT} & 95.75\% & 65.66\% & 74.65\% & 69.89\% & 82.15\% & 63.58\% & 25.84 \\
{RF} & 98.34\% & 83.11\% & 87.05\% & 85.02\% & 91.26\% & 82.76\% & 806.07 \\
{MLP} & 90.49\% & 25.35\% & 21.71\% & 23.46\% & 59.81\% & 24.55\% & 74.06 \\
{XGB} & 97.22\% & 59.46\% & 89.12\% & 71.28\% & 79.50\% & 58.84\% & 3.36 \\
\toprule
\end{tabular*}
\end{table*}

Table ~\ref{tab8} shows the performance of models in a parallel approach after performing oversampling. This table also contains the runtime.

\begin{table*}
\caption{Classification models' performance after oversampling technique\label{tab8}}
\begin{tabular*}{\textwidth}{@{\extracolsep{\fill}}c c c c c c c c@{}}
\toprule
\textit{ML Model} & \textit{Accuracy} & \textit{Recall} & \textit{Precision} & \textit{F1 Score} & \textit{ROC AUC} & \textit{FPR} & \textit{Run Time} \\
\hline
{LR} & 92.48\% & 68.68\% & 40.73\% & 51.11\% & 81.28\% & 67.22\% & 8.71\\
{DT} & 96.72\% & 59.52\% & 78.50\% & 67.78\% & 79.25\% & 58.58\% & 50.48\\
{RF} & 98.91\% & 94.38\% & 85.95\% & 89.93\% & 96.78\% & 94.10\% & 1379.10\\
{MLP} & 91.71\% & 20.14\% & 19.60\% & 19.96\% & 57.52\% & 19.42\% & 76.55\\
{XGB} & 98.21\% & 81.40\% & 86.20\% & 83.79\% & 90.36\% & 80.46\% & 5.08\\
\toprule
\end{tabular*}
\end{table*}

Figure 5 also shows the performance of the classification models on a bar graph.

\begin{figure}[ht]
    \centering
    \includegraphics[width=\columnwidth]{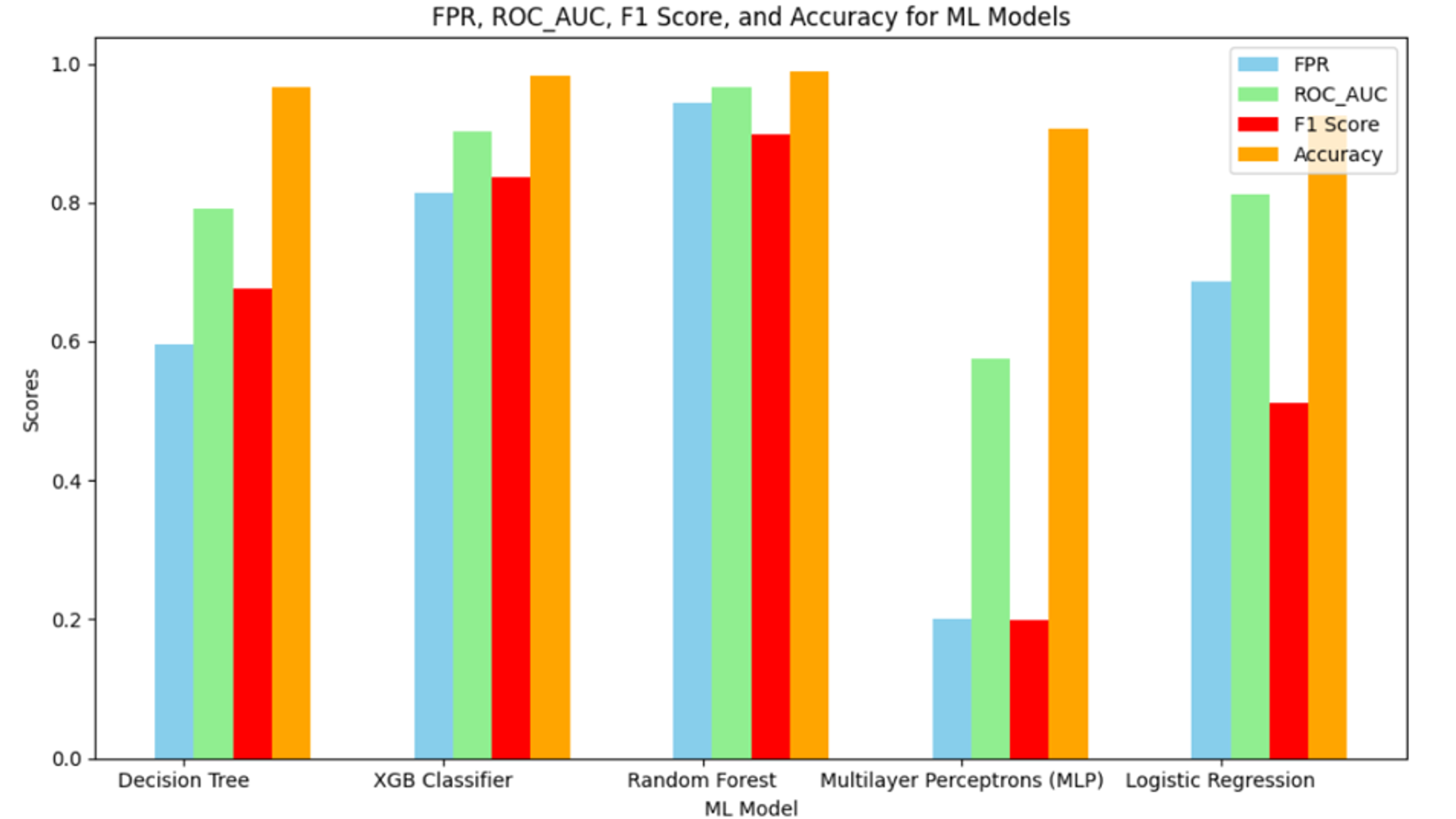}
    \caption{Bar chart of classification models’ performance after oversampling}
    \label{fig5}
\end{figure}

\section{Discussion}
\label{sec 5}

Comparison of the results in Tables ~\ref{tab7} and ~\ref{tab8} leads to valuable conclusions. RF shows the best performance under both conditions with a recall of 83.1\% on an unbalanced data set and 94.3\% for a data set balanced using an oversampling method. However, the second and third places differ for balanced and unbalanced data sets. According to RF, DT performs better than other models in an unbalanced set (recall = 65.6\%), while XGB shows a good performance of (81.3\%) in a balanced set. Furthermore, MLP performs poorly under both conditions, with recall below 30\%, while its running time is second only to RF, showing that this model is very slow and inefficient in detecting attacks. 

Overall, it is evident that oversampling (SMOTE) significantly increases the performance of the models, especially for RF and XGB, which are ensemble algorithms. However, the important point that should not be overlooked is the incredibly long run time of RF. It took 15 to 25 minutes to run RF on the test device, which is significant for use in cybersecurity systems. Systems of this type are designed to detect attacks in near real time. Therefore, it makes no sense to equip them with slow classifiers. XGB, on the other hand, required the shortest execution time in the test set (less than 10 seconds) with relatively high recall and precision. This shows that XGB can be chosen as the top model due to its extremely fast and precise performance.

\newpage

\section{Conclusions}
\label{sec 6}

This study examined the performance of five different classification models in determining zero-day attacks on a highly imbalanced data set. What sets this research apart is the following.

\begin{enumerate}
\item	We designed the framework to include zero-day attacks in the test set without biasing the model to the order of the included samples. 

\item   Despite other studies that examine ML models based only on performance metrics, our framework also considered the running time of the models in addition to the performance metrics.

\item   We have shown the positive impact of oversampling on the metrics of machine learning models on imbalanced data compared to its absence 

\item   Our analysis found that despite the time-consuming use of grid search, this cost does not significantly impact model training time as it is mainly used in earlier stages of the process.
\end{enumerate}

The zero-day attacks selected for this study are the 4 least common attacks in the applied data set, which are isolated by the classifiers during the training phase. Therefore, ML models should classify them as unseen samples only in the test set, similar to what happens with network intrusions. Evaluating the performance metrics and running time of the models shows that RF performs best regardless of whether oversampling was used or not to balance the data set. However, it cannot be crowned a champion model due to its significantly long running time. XGB has the lowest running time in the oversampled data set and is second in performance after RF, which means that XGB is the best candidate for use in cyberattack detection systems, among others.

\section*{Declarations}
All authors declare that they have no conflicts of interest.

\bibliographystyle{unsrt}
\bibliography{references}  

@article{ref-journal1,
  title={A Lightweight Phishing Detection System Based on Machine Learning and URL Features},
  author={Belfedhal, A.E. and Belfedhal, M.A.},
  journal={International Conference on Managing Business Through Web Analytics},
  pages={307--319},
  year={2022}
}

@article{ref-journal2,
  title={Detection of phishing websites using an efficient machine learning framework},
  author={Kumar, D.N. and Hemanth, N.S.R. and Premnath, S. and Kumar, V.N. and Uma, S.},
  journal={International Journal of Engineering Research and Technology},
  volume={9},
  number={5},
  pages={1282--1286},
  year={2020}
}

@article{ref-journal3,
  title={A review of Machine Learning-based zero-day attack detection: Challenges and future directions},
  author={Guo, Y.},
  journal={Computer Communications},
  volume={198},
  pages={175--185},
  year={2023}
}

@article{ref-journal4,
  title={Deep character-level anomaly detection based on a convolutional autoencoder for zero-day phishing URL detection},
  author={Bu, S.J. and Cho, S.B.},
  journal={Electronics},
  volume={10},
  pages={1492},
  year={2021}
}

@article{ref-journal5,
  title={Detecting web attacks using random undersampling and ensemble learners},
  author={Zuech, R. and Hancock, J. and Khoshgoftaar, T.M.},
  journal={Journal of Big Data},
  volume={8},
  number={1},
  pages={75},
  year={2021}
}

@article{ref-journal6,
  title={Towards benchmark data sets for machine learning based website phishing detection: An experimental study},
  author={Hannousse, A. and Yahiouche, S.},
  journal={Engineering Applications of Artificial Intelligence},
  volume={104},
  pages={104347},
  year={2021}
}

@article{ref-journal7,
  title={Explainable machine learning for phishing feature detection},
  author={Calzarossa, M.C. and Giudici, P. and Zieni, R.},
  journal={Quality and Reliability Engineering International},
  volume={40},
  number={1},
  pages={362--373},
  year={2024}
}

@article{ref-journal8,
  title={From zero-shot machine learning to zero-day attack detection},
  author={Sarhan, M. and Layeghy, S. and Gallagher, M. and Portmann, M.},
  journal={International Journal of Information Security},
  volume={22},
  number={4},
  pages={947--959},
  year={2023}
}

@article{ref-journal9,
  title={A comparison study of generative adversarial network architectures for malicious cyber-attack data generation},
  author={Peppes, N. and Alexakis, T. and Demestichas, K. and Adamopoulou, E.},
  journal={Applied Sciences},
  volume={13},
  number={12},
  pages={7106},
  year={2023}
}

@article{ref-journal10,
  title={Towards a standard feature set for network intrusion detection system data sets},
  author={Sarhan, M. and Layeghy, S. and Portmann, M.},
  journal={Mobile Networks and Applications},
  pages={1--14},
  year={2022}
}

@article{ref-journal11,
  title={Visualphishnet: Combining long-term recurrent convolutional and graph convolutional networks to detect phishing sites using URL and HTML},
  author={Ariyadasa, S. and Fernando, S. and Fernando, S.},
  journal={IEEE Access},
  volume={8},
  pages={82355--82375},
  year={2022}
}

@article{ref-journal12,
  title={On the analysis of open source data sets: validating IDS implementation for well-known and zero day attack detection},
  author={Serinelli, B.M. and Collen, A. and Nijdam, N.A.},
  journal={Procedia Computer Science},
  volume={191},
  pages={192--199},
  year={2021}
}

@article{ref-journal13,
  title={Accurate and fast URL phishing detector: a convolutional neural network approach},
  author={Wei, W. and Ke, Q. and Nowak, J. and Korytkowski, M. and Scherer, R. and Wo{\'z}niak, M.},
  journal={Computer Networks},
  volume={178},
  pages={107275},
  year={2020}
}

@inproceedings{ref-journal14,
  title={Visualphishnet: Zero-day phishing website detection by visual similarity},
  author={Abdelnabi, S. and Krombholz, K. and Fritz, M.},
  booktitle={Proceedings of the 2020 ACM SIGSAC Conference on Computer and Communications Security},
  pages={1681--1698},
  year={2020}
}

@inproceedings{ref-journal15,
  title={Towards the detection of malicious URL and domain names using machine learning},
  author={Ghalati, N.F. and Ghalaty, N.F. and Barata, J.},
  booktitle={Technological Innovation for Life Improvement: 11th IFIP WG 5.5/SOCOLNET Advanced Doctoral Conference on Computing, Electrical and Industrial Systems (DoCEIS 2020)},
  pages={109--117},
  year={2020}
}

@article{ref-journal16,
  title={Machine learning based phishing detection from URLs},
  author={Sahingoz, O.K. and Buber, E. and Demir, O. and Diri, B.},
  journal={Expert Systems with Applications},
  volume={117},
  pages={345--357},
  year={2019}
}

@article{ref-journal17,
  title={An overview of principal component analysis},
  author={Karamizadeh, S. and Abdullah, S.M. and Manaf, A.A. and Zamani, M. and Hooman, A.},
  journal={Journal of Signal and Information Processing},
  volume={4},
  number={3},
  pages={173--175},
  year={2013}
}

@book{ref-book1,
  title={Principal Component Analysis: Engineering Applications},
  editor={Sanguansat, P.},
  publisher={Publishing House},
  address={Rijeka, Croatia},
  pages={132--158},
  year={2012}
}

@article{ref-journal19,
  title={Two novel SMOTE methods for solving imbalanced classification problems},
  author={Bao, Y. and Yang, S.},
  journal={IEEE Access},
  volume={11},
  pages={5816--5823},
  year={2023}
}

@article{ref-journal20,
  title={Hyper-parameter optimization: A review of algorithms and applications},
  author={Yu, T. and Zhu, H.},
  journal={arXiv preprint arXiv:2003.05689},
  year={2020}
}

@article{ref-journal21,
  title={Performance Comparison of Grid Search and Random Search Methods for Hyperparameter Tuning in Extreme Gradient Boosting Algorithm to Predict Chronic Kidney Failure},
  author={Anggoro, D.A. and Mukti, S.S.},
  journal={International Journal of Intelligent Engineering and Systems},
  volume={14},
  number={6},
  year={2021}
}

@article{ref-journal22,
  title={Grid search in hyperparameter optimization of machine learning models for prediction of HIV/AIDS test results},
  author={Belete, D.M. and Huchaiah, M.D.},
  journal={International Journal of Computers and Applications},
  volume={9},
  pages={875--886},
  year={2022}
}

@article{ref-journal23,
  title={Decision forest: Twenty years of research},
  author={Rokach, L.},
  journal={IEEE Access},
  volume={27},
  pages={111--125},
  year={2016}
}

@article{ref-journal24,
  title={Random forests},
  author={Breiman, L.},
  journal={Machine Learning},
  volume={45},
  pages={5--32},
  year={2001}
}

@inproceedings{ref-journal25,
  title={XGBoost: A scalable tree boosting system},
  author={Chen, T. and Guestrin, C.},
  booktitle={Proceedings of the 22nd ACM SIGKDD International Conference on Knowledge Discovery and Data Mining},
  pages={785--794},
  year={2016}
}

@article{ref-journal26,
  title={Cardiac arrhythmia classification by multi-layer perceptron and convolution neural networks},
  author={Savalia, S. and Emamian, V.},
  journal={Bioengineering},
  volume={5},
  number={2},
  pages={35},
  year={2018}
}

@article{ref-journal27,
  title={On evaluation metrics for medical applications of artificial intelligence},
  author={Hicks, S.A. and Strümke, I. and Thambawita, V. and Hammou, M. and Riegler, M.A. and Halvorsen, P. and Parasa, S.},
  journal={Scientific Reports},
  volume={12},
  number={1},
  pages={5979},
  year={2022}
}

@article{ref-journal28,
  title={Magician's corner: 9. Performance metrics for machine learning models},
  author={Erickson, B.J. and Kitamura, F.},
  journal={Radiology: Artificial Intelligence},
  volume={3},
  pages={200126},
  year={2021}
}






\end{document}